%% file: main.tex
\documentclass{trbunofficial}
\usepackage{graphicx}
\usepackage{amsmath}

\usepackage[hidelinks]{hyperref}
\usepackage[T1]{fontenc}
\usepackage{newtxtext,newtxmath} 

\AuthorHeaders{Kuncheria, Walker and Macfarlane}
\title{ Beyond Centrality: Understanding Urban Street Network Typologies Through Intersection Patterns}

\author{
  \textbf{Anu Kuncheria, Corresponding Author} \\
  anu\_kuncheria@berkeley.edu\\
  Department of Civil and Environmental Engineering\\
  University of California, Berkeley\\
  \textbf{Joan L. Walker}\\
 joanwalker@berkeley.edu\\
 Department of Civil and Environmental Engineering\\
  University of California, Berkeley\\
 \textbf{Jane Macfarlane}\\
 janemacfarlane@berkeley.edu\\
 Smart Cities Research Center\\
 University of California, Berkeley
}

\begin{document}
\input{author}

\section{Abstract}
The structure of road networks plays a pivotal role in shaping transportation dynamics. It also provides insights into how drivers experience city streets and help uncover each urban environment's unique characteristics and challenges. Consequently, characterizing cities based on their road network patterns can facilitate the identification of similarities and differences, informing collaborative traffic management strategies, particularly at a regional scale. While previous studies have investigated global network patterns for cities, they have often overlooked detailed characterizations within a single large urban region. Additionally, most existing research uses metrics like degree, centrality, orientation, etc., and misses the nuances of street networks at the intersection level, specifically the geometric angles formed by links at intersections, which could offer a more refined feature for characterization.  To address these gaps, this study examines over 100 cities in the San Francisco Bay Area. We introduce a novel metric for classifying intersections, distinguishing between different types of 3-way and 4-way intersections based on the angles formed at the intersections. Through the application of clustering algorithms in machine learning, we have identified three distinct typologies—grid, orthogonal, and organic cities—within the San Francisco Bay Area. We demonstrate the effectiveness of the metric in capturing the differences between cities based on street and intersection patterns. The typologies generated in this study could offer valuable support for city planners and policymakers in crafting a range of practical strategies tailored to the complexities of each city’s road network, covering aspects such as evacuation plans, traffic signage placements, and traffic signal control.


\hfill\break%
\noindent\textit{Keywords}: Street networks, Geospatial data, Cities, Machine learning, Urban analytics, City science
\newpage

\section{Introduction}
The road network serves as the backbone of any city, providing essential connections between different parts of the urban landscape.  The structure and connectivity of these roads influence how drivers experience the city as they navigate streets and intersections, thereby affecting traffic flow, behavior, and the overall dynamics of the city. These networks are carefully planned in some cities, while they have evolved organically in others \cite{barthelemy_self-organization_2013, batty_thinking_2017}. In planned cities, drivers benefit from predictable patterns, clear signage, and consistent traffic control measures, which lead to smoother navigation. In contrast, cities with non-grid road networks may present more challenges, such as irregular street layouts and unexpected intersections. 

Analyzing cities based on their street layouts offers valuable insights for formulating effective traffic management strategies. By characterizing each city, it becomes feasible to devise efficient and attuned strategies to each layout's distinctive features, thereby optimizing the overall functionality of the urban landscape. Further, this understanding can foster collaboration among cities within a broader urban region.
 
Over the last two decades, there has been a substantial body of research in network science, with a focus on developing metrics to delineate and understand networks. In the realm of transportation networks, there has been an emphasis on classifying cities based on both network topology and geometric features \cite{strano_urban_2013,boeing_urban_2019,crucitti_centrality_2006,badhrudeen_geometric_2022}. Despite the extensive exploration of topological metrics, there is still a gap in investigating geometric metrics. Specifically, no studies have examined the identification of intersection patterns based on the geometric angles formed. These intersection patterns, especially at 3-way and 4-way intersections, exhibit variations contributing to distinct intersection configurations, thereby resulting in different road network structures. Explicitly considering these intersection patterns promises to generate better, more accurate, and representative typologies. 

Furthermore, the majority of existing literature is dedicated to characterizing cities worldwide, enabling the recognition of overarching trends and patterns. However, to derive practical insights and effectively apply them to planning and design, performing clustering at a metropolitan level or contiguous regional level is crucial.

To address these research gaps, our study concentrates on characterizing cities in the expansive urban region of the San Francisco Bay Area, encompassing over 90 cities. Through the application of various topological and geometric measures, we establish distinct typologies for cities based on their network structure. Additionally, we introduce a novel metric to identify node patterns linked to three-way and four-way intersection types. Subsequently, we develop two clustering models: a baseline employing existing metrics and an enhanced model integrating additional measures. Our analysis entails a thorough comparison and contrast of cities within these two models, emphasizing the effectiveness of the new metric in delineating city structures. Lastly, we engage in a discussion on how these classifications can inform effective transportation management strategies.

The subsequent sections of this paper are structured as follows: Section 2 presents a comprehensive overview of previous studies on city characterization. In Section 3, we detail the study's data and methods. Section 4 discusses the results, and Section 5 provides a discussion. Finally, the conclusions are presented in Section 6. 

\section{Literature Review}
Several studies have explored urban street network layouts, utilizing both topological and geometrical metrics to comprehensively characterize the overall structure \cite{ lin_comparative_2017,kirkley_betweenness_2018,shang_statistical_2020,xie_measuring_2005, cardillo_structural_2006,masucci_exploring_2014, wang_improved_2017,akbarzadeh_role_2019, fancello_modeling_2014,jiang_topological_2004,boeing_urban_2019, tsiotas_topology_2017, boeing_morphology_2019, barthelemy_betweenness_2004,crucitti_centrality_2006,strano_urban_2013,badhrudeen_geometric_2022}. Topological metrics unveil connectivity patterns, while geometrical metrics elucidate spatial features within the network.

Past research has characterized cities worldwide using a combination of topological and geometric network properties, with the aim of discerning similarities and differences among them. For instance, \citep{boeing_urban_2019} classified 100 cities worldwide into three primary and eight secondary levels based on four features of average node degree, orientation order, median street length, and average street circuit. They introduced a novel metric called orientation order to gauge the overall alignment of streets within a city. Their metrics were designed to assess whether a city's street layout resembled a grid, and they observed that cities in the United States exhibited a stronger grid-like pattern compared to those in other parts of the world. \citep{crucitti_centrality_2006} utilized four node centrality metrics: closeness centrality, betweenness centrality, straightness centrality, and information centrality, to classify 18 cities into three types: planned, self-organized, and model. They demonstrated that employing various centralities enables the capture of valuable structural properties within networks. \citep{strano_urban_2013} examined the geometric properties such as average street length, the distribution of angles, and the proportion of dead ends. Additionally, it explored four centrality measures, categorizing cities into two groups based on the presence or absence of significant geographical constraints. In their study, \citep{badhrudeen_geometric_2022} classified 80 cities into five distinct categories: Gridiron, Long Link, Organic, Hybrid, and Mixed cities, by examining the topological and geometric characteristics of road networks. This classification utilized various metrics, including node degree distribution, intersection angle distribution, and link length distribution

In addition to city clustering, a significant body of literature is dedicated to understanding network structure properties using various metrics \cite{lin_comparative_2017, shang_statistical_2020, zhang_backbone_2017,cardillo_structural_2006,masucci_exploring_2014, wang_improved_2017}. \citep{xie_measuring_2007} presented three new metrics aimed at capturing network heterogeneity, including entropy, connection patterns, and continuity. Their study demonstrated the utility of these measures in quantifying and comparing structural attributes of road networks. \citep{buhl_topological_2006} investigated topological metrics, network efficiency, and network robustness as means to characterize the properties of street networks. \citep{chan_urban_2011} analyzed the distributions of various geometric metrics of street links, including link length, link angle, and double-angle, across 20 German cities. \citep{jiang_topological_2007} examined the topological metrics such as degree, path length, and clustering coefficient across 40 cities, illustrating their small-world properties with scale-free characteristics. Their findings showed that roughly 80\% of streets have lengths below the average, while 20\% have lengths that exceed the average.

While both topological and geometric features have been extensively investigated, a notable gap exists in capturing the nuanced geometric angles of the network. Two studies have captured the overall street angles \cite{strano_urban_2013,badhrudeen_geometric_2022}; however, its approach treats all angles independently, neglecting the intricate patterns formed at intersections as a collective entity. This level of detail is essential for determining whether a city can be classified as having a grid-like structure or not. The precise capture of intersection patterns based on the angles formed is paramount for the development of accurate network typologies. Furthermore, the majority of existing literature focuses on cities globally, allowing for the identification of broad trends. However, it's crucial to conduct clustering analysis at a regional level to transform findings into practical strategies and align them with planning and design processes. This study addresses this need by concentrating on California's San Francisco Bay Area region.

\section {Methods}
\subsection {Data}
Our study focuses on the San Francisco Bay Area in California, encompassing nine counties and 101 municipalities. In California, municipalities have the flexibility to use either the term "city" or "town" in their official names, as there is no legal differentiation between the two \cite{california-law}. To ensure a fair and balanced comparison, we have excluded very small cities with sparse road networks and only included municipalities with populations exceeding 5000, resulting in a total of 94 cities under consideration.

The road network used in this study is the Mobiliti Bay Area network \cite{chan_simulating_2022}, which serves as the foundation for an urban-scale, parallel discrete event simulator jointly developed by the Lawrence Berkeley National Laboratory (LBNL) and the Smart Cities Research Center at the Institute for Transportation Studies at UC Berkeley. This road network graph is derived from a professional HERE Technologies map \cite{here_tech}. The Mobiliti map is designed as a directional map, with links having a start and end node, representing the direction of traffic flow. The map was pre-processed to transform it into a primal graph representation, where nodes represent intersections or dead ends, and links denote streets connecting these nodes. The graph is then clipped to each city's boundaries using administrative boundary data from the Metropolitan Transportation Commission (MTC) \cite{mtc}.

\subsection {Existing Metrics}
As noted in the previous section, metrics used to quantify urban street networks encompass both topological and geometrical aspects. In quantifying the topological properties of road networks, the literature explores various measures, including degree, betweenness centrality, closeness centrality, and clustering coefficient \cite{lin_comparative_2017, kirkley_betweenness_2018}. Each metric captures different aspects of road networks: degree measures connectivity at intersections, betweenness centrality assesses the network's ability to facilitate paths between regions, and closeness centrality indicates proximity within the network. For our study, we specifically focus on node degree and betweenness centrality as topological metrics, as they provide comprehensive insights into the overall properties of road networks.

The degree of a node reflects the number of connections it possesses with other nodes, identifying the most interconnected nodes within a city (Equation \ref{eq:degree}). In a directed graph G with N nodes and E edges, two types of degrees exist: in-degree, indicating connections into a node, and out-degree, representing connections exiting the node. Generally, in-degrees and out-degrees are equal, except for intersections with one-way streets. For instance, if a one-way street leads into an intersection, the in-degree of that intersection would exceed its out-degree.

\begin{equation}
D(i) =  \sum_{j \subseteq N} a_{ij}
\label{eq:degree}
\end{equation}

where: \\
\indent \indent  $D(i)$ is the degree of node $i$\\
\indent \indent  $a_{ij}$ is the element of the adjacency matrix, when a node i is connected to the node j,\\
\indent \indent $a_{ij}$ = 1, otherwise $a_{ij}$ = 0. \\

\begin{figure}[h]
  \includegraphics[width= 1.05\columnwidth]
  {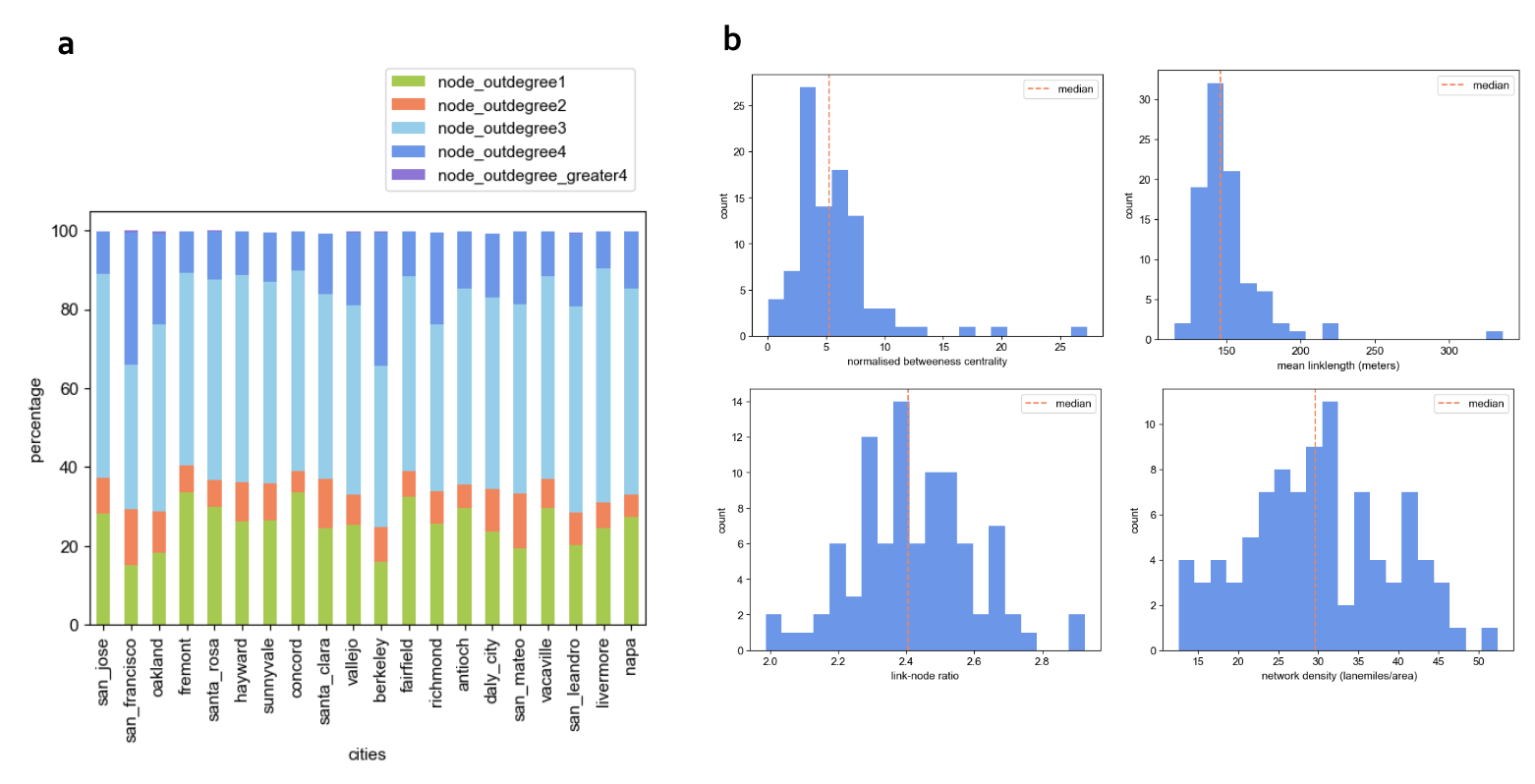}
  \caption{ a. The out-degree distribution of the top twenty cities in the Bay Area. b. Distribution of various geometric metrics across cities.}
  \label{fig:existing_met_dist}
\end{figure}

For all 94 cities in our analysis, both in-degree and out-degree range from 1 to 6. Most nodes exhibit equal in-degree and out-degree. The median percentage of nodes with different in and out degrees is 3\%. For the clustering in the next step, we incorporate the proportion of nodes falling into 5 node degrees (1, 2, 3, 4, and more than 4) as a feature. Figure \ref{fig:existing_met_dist}a illustrates the distribution of node degree proportions in the top twenty cities in the Bay Area.

Betweenness centrality (BC) evaluates the degree to which a node lies on the shortest paths connecting pairs of other nodes, measuring the node's intermediate importance in facilitating interactions among other nodes \cite{kirkley_betweenness_2018, lin_comparative_2017}. Nodes with high betweenness centrality are pivotal for network resilience, as they serve as critical connectors between multiple areas within a region. It is calculated as shown in  Equation \ref{eq:bc}. To ensure equitable comparisons across cities of varying sizes, the BC is normalized by the number of nodes in a city. The study employs the median values of the normalized BC for analysis.

\begin{equation}
BC(i) = \frac{1}{c}  \sum_{a \neq b \subseteq N} \frac{\sigma_{ab(i)}}{\sigma_{ab}}
\label{eq:bc}
\end{equation}

where: \\
\indent \indent $BC(i)$ is the BC of node $i$\\
\indent \indent $\sigma_{ab}$ is the number of shortest paths going from nodes $a$ to $b$\\
\indent \indent $\sigma_{ab(i)}$ is the number of the shortest paths going from $a$ to $b$ through node $i$\\
\indent \indent $c$ is a normalisation constant\\

For geometric metrics, we computed the mean link length, network density, and link-node ratio for all cities. The distributions for the cities are shown in Figure \ref{fig:existing_met_dist}b. Additionally, link bearings for each link in a city were calculated \cite{noauthor_osmnx_2016} and discretized into 20-degree bins. The proportion of links in each bin is considered a feature for the subsequent analysis. An adjustment was made to the order of the bins, ensuring that the first bin with a large proportion is considered bin 1. This adjustment is implemented so that the clustering is not influenced by specific cardinal directions (e.g., east-west or north-south), as the primary focus is on identifying the number of predominant orientation directions rather than the specific orientation itself.  Additionally, we use the number of dominant bins as another feature for analysis. A dominant bin is defined as a bin where more than 10 \% of a link's orientation falls. If a city's streets are oriented predominantly in a few directions, the number of dominant bins will be higher, as more bins will satisfy the 10\% threshold. Conversely, if a city's streets are oriented in many directions, there won't be any bins that pass the 10\% threshold. In addition to the bearing proportion, this metric summarizes a city's overall orientation of streets.


\subsection {Proposed Metric}
To capture the nuances of the geometry of the road network, we propose a novel metric for identifying intersection patterns within the network. These patterns are discerned by analyzing the angles created by outgoing links at every intersection for nodes with degrees 3 and 4. The reason for choosing out-degree instead of in-degree for node patterns is that we want to capture the angles the vehicles have to take when they arrive at intersections. By capturing the angles of outgoing links, the metric aims to provide a more nuanced understanding of how vehicles navigate intersections. The angles formed by links at intersections are then categorized as acute, obtuse, right or reflex angles, which are then used in conjunction with node degrees to facilitate pattern recognition. 

 Since in our graph, nodes correspond to intersections, and links represent edges connecting these nodes, the curvature of the link is not captured in the graph. To address this, we incorporate the shape points provided in the links file. The shape points of a link represents the curvature or sharp bend in the link, reflecting real-world features. Thus, by calculating the angle formed by the start node and the first shape point of a link, we will be able to capture the true geometry of the links meeting at the intersection. The angles between links are calculated based on coordinate geometry  (Equation \ref{eq:angle_calculate}). Given three points, denoted as a, o, and b, formed by two lines, we can calculate the arctangent to determine the angles formed by the line segments connecting o to a and o to b. Subsequently, we compute the difference between these angles and convert the result from radians to degrees. This process is repeated for all angles formed at the intersection.

\begin{equation}
  \text{angle}(a,o,b) =
    \begin{cases}
         \text{degrees}(\text{atan2}(b_y - o_y, b_x - o_x) - \text{atan2}(a_y - o_y, a_x - o_x)) & \text{ if angle}(a,o,b) > 0 \\

         \text{degrees}(\text{atan2}(b_y - o_y, b_x - o_x) - \text{atan2}(a_y - o_y, a_x - o_x)) + 360  & \text{ if angle}(a,o,b) < 0 \\
         
    \end{cases}
    \label{eq:angle_calculate}
\end{equation}

where: \\
\indent \indent $a_x, a_y$ are the coordinates of point $a$, \\
\indent \indent $o_x, o_y$ are the coordinates of point $o$, \\
\indent \indent $b_x, b_y$ are the coordinates of point $b$ \\
\indent \indent $degrees$ represents the conversion of an angle from radians to degrees

\begin{figure}[h]
  \includegraphics[width=.5\columnwidth]
    {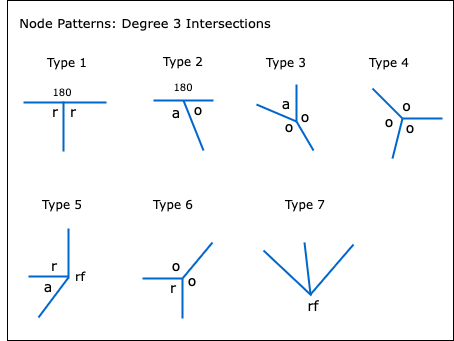}
  \includegraphics[width=.5\columnwidth]
    {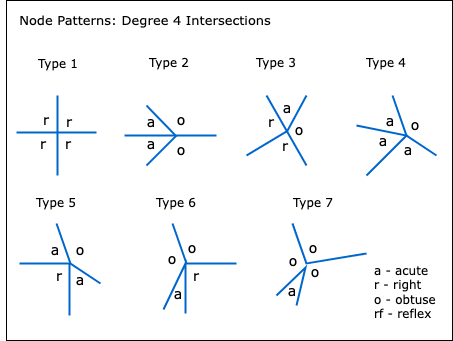}
  \caption{3-way (left) and 4-way intersection patterns (right) for nodes.}
  \label{fig:nodepatterns}
\end{figure}

Once the angles are calculated, patterns are identified for each node degree. For nodes with a degree of 3 (three-way intersections), seven distinct patterns emerge from various angle configurations. For instance, in degree 3 Type 1, a traditional T intersection is represented with two right angles and one 180-degree angle. When the angles comprise acute, obtuse, and 180-degree angles, it falls under Type 2. Additional types demonstrate alternative three-way intersection configurations \ref{fig:nodepatterns}.

Similarly, for nodes with a degree of 4 (four-way intersections), seven distinct patterns emerge based on the angles formed by the four links. Type 1 signifies a scenario where all four angles are perfect right angles, commonly found in planned grid layouts.  The other types are as shown in Figure \ref{fig:nodepatterns}. 

Examining node patterns across the entire Bay Area, for degree 3 intersections, the most common type is Type 1 (74\%), followed by Type 2 (13\%), and then Type 6 (6\%). In the case of 4-way intersections, the predominant pattern is Type 1 (69\%), with Types 3 (18\%) and 2 (6\%) following in frequency.

\begin{figure}[h]
  \includegraphics[width= 1\columnwidth]
{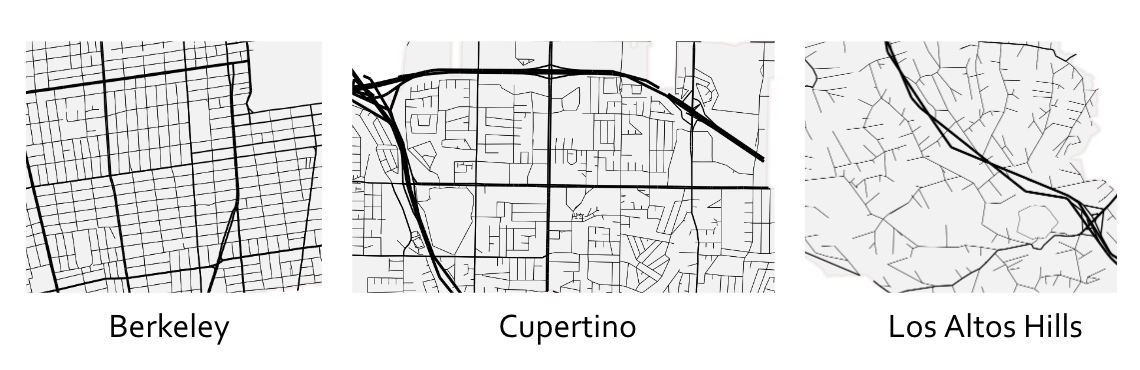}
  \caption{ The figure displays a zoomed-in portion of the road networks for selected cities.}
  \label{fig:example_metric}
\end{figure}

To illustrate distinctive node patterns, we examine three cities with notable differences in their road networks (Figure \ref{fig:example_metric}). Berkeley, known for its planned grid network, displays a substantial concentration of both degree 4 and 3 intersections. Within degree 4 intersections, Berkeley stands out with the highest proportion of Type 1 intersections, emphasizing the city's grid network layout. Within degree 3 intersections, the majority are Type 1 T intersections.

In Cupertino, degree 3 intersections dominate, closely followed by degree 1 intersections. Notably, Cupertino's degree 3 intersections are predominantly characterized by Type 1 intersections, classic T intersections. In contrast to Berkeley, Cupertino showcases a mix of Type 1 and Type 3 intersections among its degree 4 intersections, presenting a mix of perpendicular and angled links. This is evident in the city's orthogonal structure and the presence of dead ends within the blocks. 

On the other hand, Los Altos Hills features a substantial proportion of degree 1 intersections, followed by degree 3 intersections. The degree 3 intersections in Los Altos Hills are primarily Type 2 and 6, featuring angled three-way intersections. For degree 4 intersections, Los Altos Hills is distinguished by a higher proportion of Type 3 intersections, again featuring angled configurations. This is evident when examining the full network of the city, where most streets wind and lack a specific order or orientation.

It is evident that the diversity in intersection types provides valuable insights into the distinct geometric patterns and urban layouts of these cities. Analyzing node patterns alongside degrees proves to be a clear and effective method for distinguishing street layouts and can be an important feature for city clustering.

\subsection{Feature Correlations}
Before clustering, we conducted a correlation analysis using the Pearson correlation coefficient. This method is commonly used to capture linear relationships between variables, enabling us to identify notable trends in our dataset. In alignment with established literature, we observed a substantial positive correlation between node in-degree and out-degree, exceeding values of 0.99 and thus providing redundant statistical information about the networks. Therefore, we use node out-degree as the degree metric for the subsequent analyses. For the remainder of this paper, when we mention "degree," it specifically refers to out-degree.

Degree 1 shows an inverse correlation with degree 4, betweenness centrality, and the link-node ratio, aligning with findings from previous studies \cite{chan_urban_2011,burghardt_road_2022,dumedah_characterising_2021}. Furthermore, it exhibits an inverse correlation with the node pattern for degree 4 Type 1. This observation aligns with the logic that cities characterized by a higher proportion of dead ends tend to have fewer 4-way intersections.

Degree 4 demonstrates a positive correlation to betweenness centrality and link-node ratio as expected. Furthermore, it displays a positive correlation with node pattern Degree 4 Type 1, indicating that cities characterized by a high proportion of 4-way intersections are more inclined to feature 90-degree angled four-ways. Moreover, the degree 4 Type 1 intersection pattern is positively correlated with the link-node ratio, expected from a gridded network structure.

Degrees 3 and 1 do not show any correlations. However, Degree 3 Type 1 is inversely correlated with Degree 1, indicating a lower proportion of T intersections associated with a city with a high proportion of dead ends. Degree 3 Type 1 also exhibits inverse correlations with Types 2, 3, and 6, which are non-right-angled 3-way intersections. 

\subsection {Clustering}

To categorize cities into distinct groups, we use unsupervised machine learning clustering algorithms. We conduct two separate clustering analyses: a baseline clustering using metrics commonly found in existing literature and an enhanced clustering where we supplement the baseline with additional metrics. This aims to highlight the distinctions between the two clustering approaches and demonstrate how the inclusion of new metrics adds value to characterizations.

In the baseline clustering, we use five metrics: node degree, betweenness centrality, mean link length, network density, and link-node ratio. For enhanced clustering, we expand this set by incorporating two additional metrics. The first addition is node patterns, the new metric introduced and explained in the previous section, represented by 14 features. The second addition is link bearings, which comprise 19 features. While \citep{boeing_urban_2019} employed orientation entropy, a derived metric from link bearings, in his study, we believe that directly including link bearings as a feature adds intrinsic value to clustering. 

We normalized all metrics to account for the different features and city sizes. Then, we conducted a factor analysis to reduce the feature size by extracting all their commonalities into a smaller set of factors. Ten factors were chosen, giving an eigenvalue greater than 1, a common standard used in the field. For clustering, we use the K-means method \cite{noauthor_k-means_2023} with varying numbers of clusters and choose the number of clusters based on the elbow method and explainability. We use the Silhouette score and the Davies-Buldin Index to evaluate clusters. The Silhouette Score quantifies how well a data point fits into its assigned cluster and how distinct it is from other clusters. Davies-Buldin Index is another metric that looks at within-cluster and between-cluster distances. It is improved (lowered) by increased separation between clusters and decreased variation within clusters.

\section {Results}
In this section, we present the results for baseline clustering, followed by enhanced clustering. Subsequently, in the next section, we offer a comparison between the two methodologies to provide insights into their differences.

\begin{figure}[ht]
  \includegraphics[width=.25\columnwidth]
    {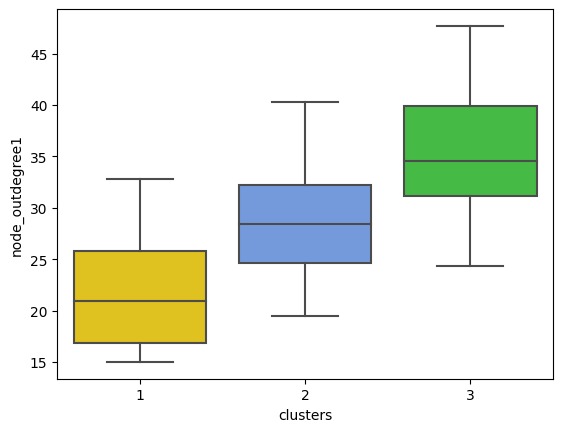}\hfill
  \includegraphics[width=.25\columnwidth]
    {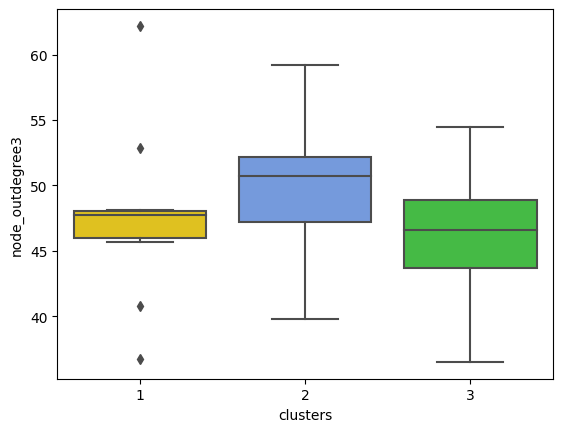}\hfill
  \includegraphics[width=.25\columnwidth]
    {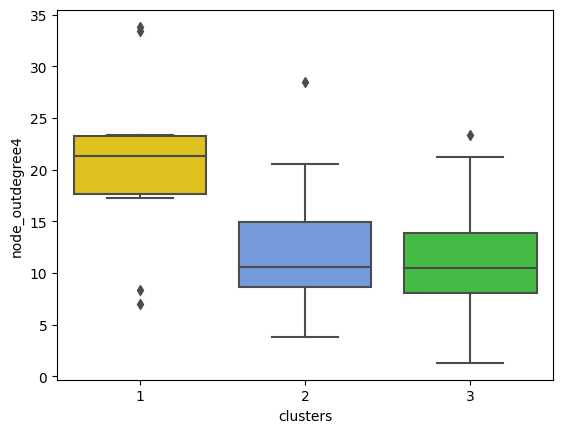}\hfill
  \includegraphics[width=.25\columnwidth]{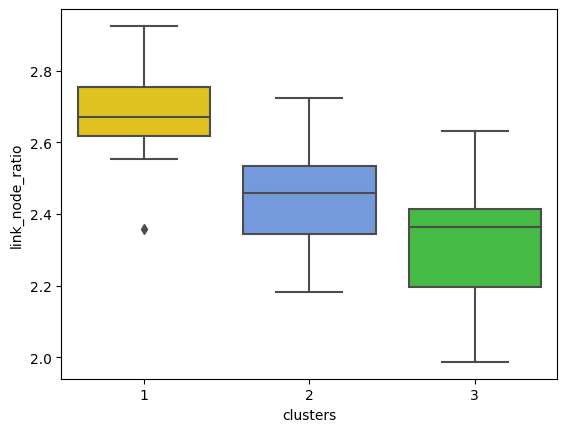}
    \includegraphics[width=.25\columnwidth]{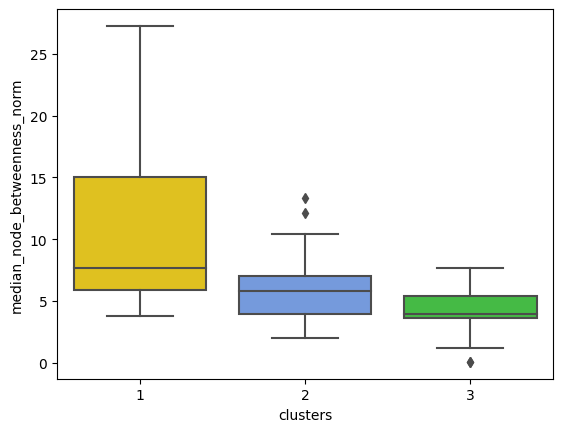}
    \includegraphics[width=.25\columnwidth]{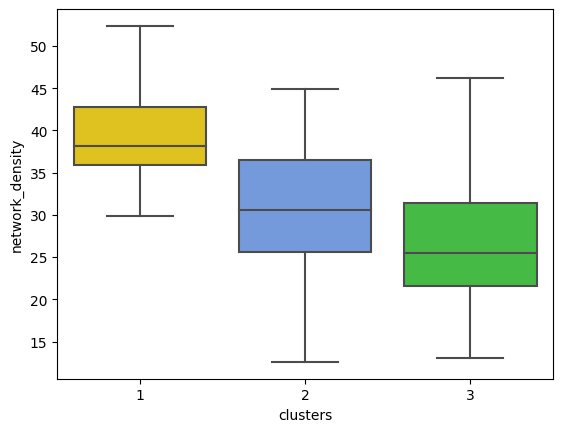}
  \caption{The figures show the distribution of metrics for baseline clusters.}
  \label{fig:baseline_metrics}
\end{figure}

In the baseline clustering, we identify three clusters, with the most crucial features being node degrees, link-node ratio, and network density (Figure \ref{fig:baseline_metrics}). Cluster 1 is characterized by high-degree 4 nodes and low-degree 1 nodes. Cluster 2 is distinguished by high-degree 3 nodes, and Cluster 3 by high-degree 1 nodes. It is commonly assumed that cities with high degree 4 are indicative of gridded urban layouts. However, upon scrutinizing city networks, it becomes apparent that several cities are misclassified. For instance, despite featuring a visible grid network, Richmond and Palo Alto are incorrectly assigned to Clusters 3 and 2, respectively. Similarly, Clusters 2 and 3, marked by a high proportion of degree 3, display numerous instances of mixed classifications. 

\begin{figure}[ht]
  \includegraphics[width=.52\columnwidth]{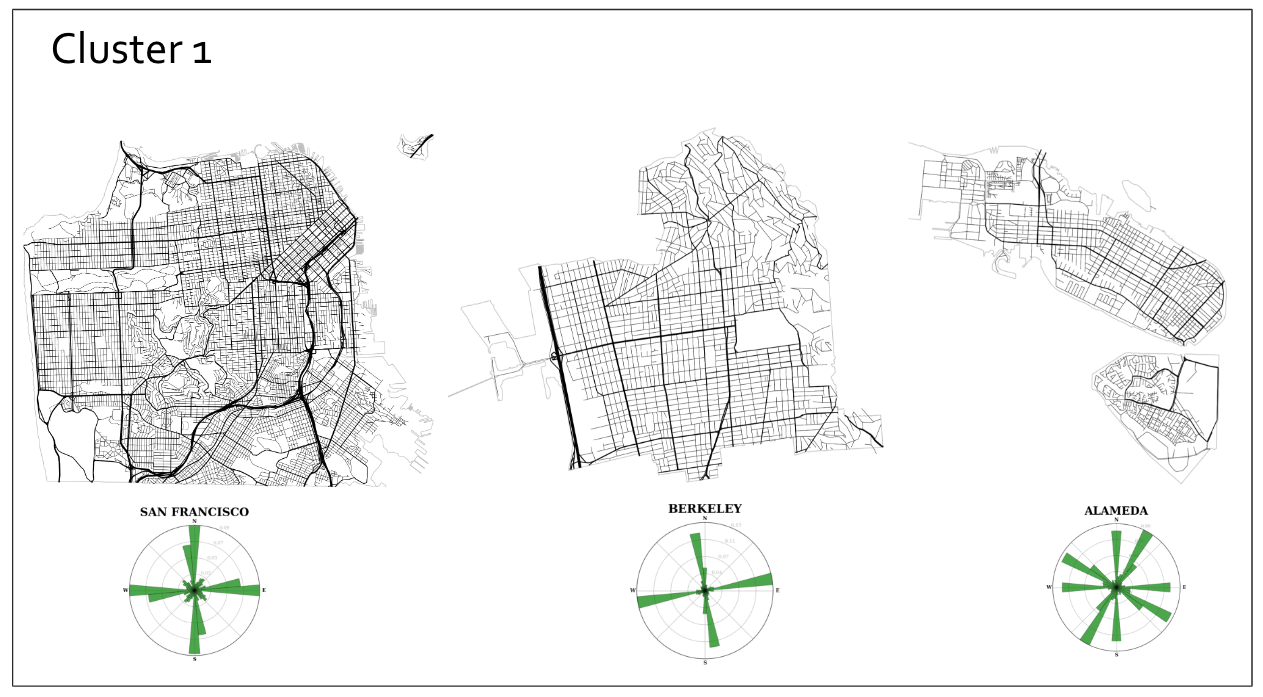}\hfill
  \includegraphics[width=.49\columnwidth]
{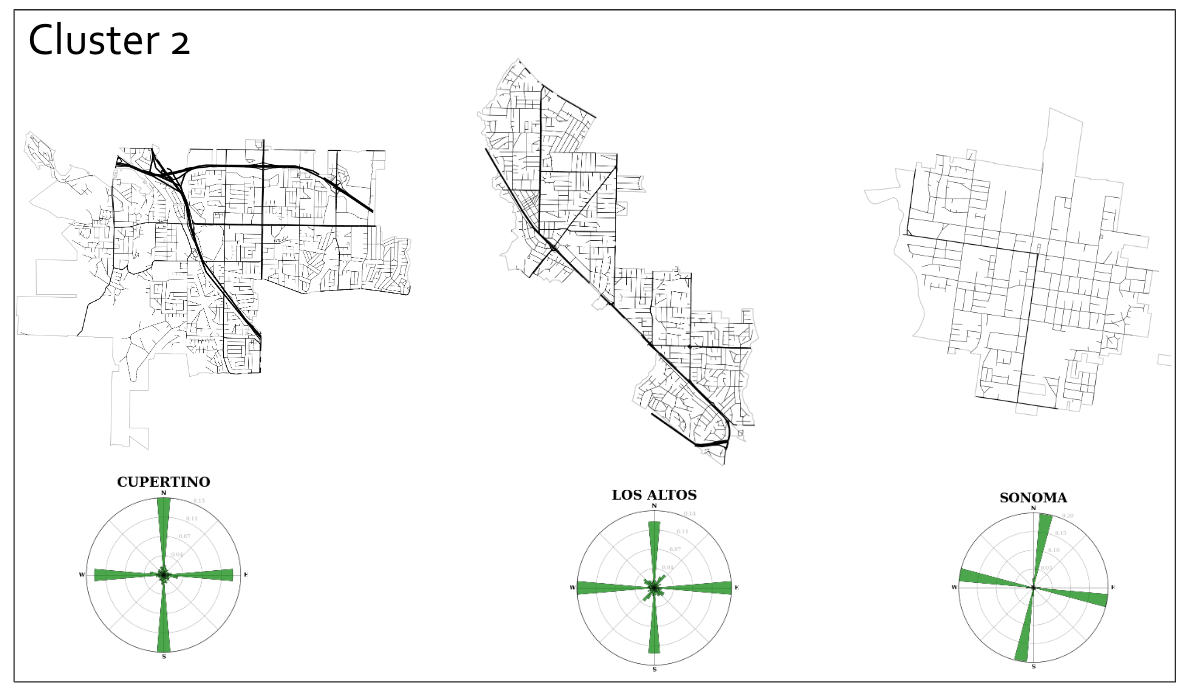}\hfill
  \includegraphics[width=.52\columnwidth]
{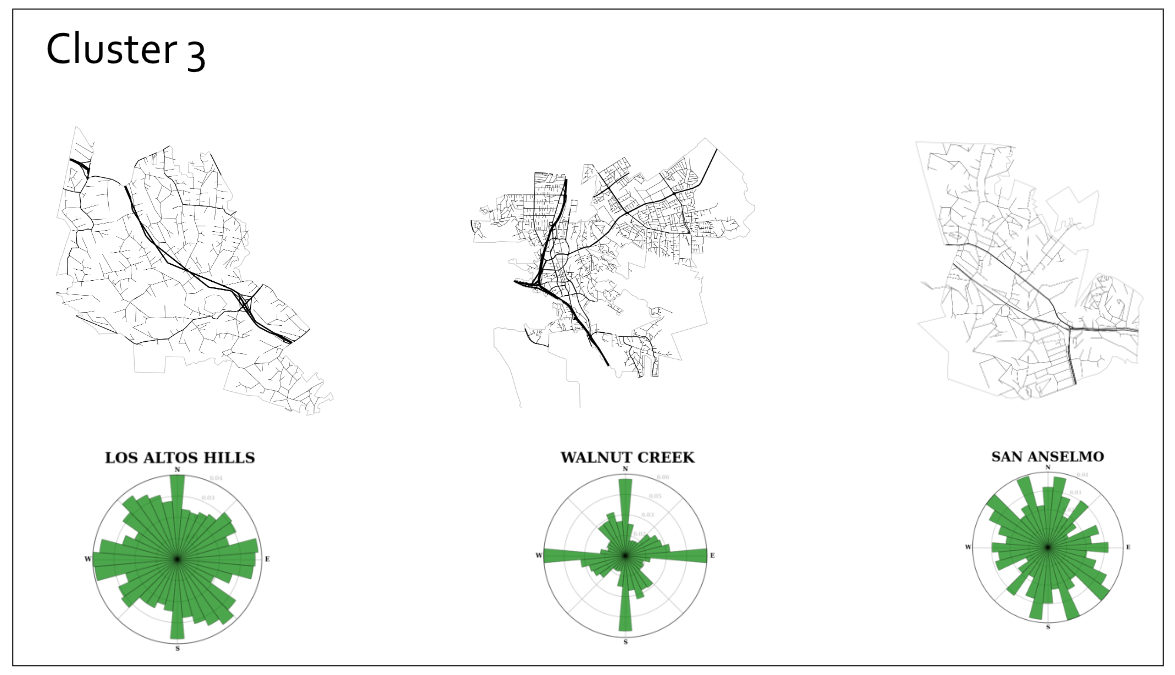}\hfill
  \caption{The figures display example cities from enhanced clustering. The green polar plots illustrate the street bearings of all links in the city.}
  \label{fig:enhanced_cities}
\end{figure}

In the enhanced clustering with added features, we obtain 3 clusters. The key features driving this classification include node degrees, node patterns, link-node ratio, and link bearings. Figure \ref{fig:enhanced_cities} shows the example cities in each cluster, and Table \ref{tab:enhanced_clusters} provides a summary for each cluster. The polar plots are generated as specified in \citep{boeing_modeling_nodate}, utilizing our road network.

Cluster 1, distinguished by a high proportion of node degree 4, a significant prevalence of Type 1 node patterns with degree 4 and 3, and a high number of dominant bearing bins, can be labelled as \textbf{Gridded cities}. These cities predominantly exhibit a grid layout with right-angled four-way intersections and three-way T intersections.

Cluster 2, featuring a high proportion of degree 3, elevated Type 1 degree 3 node patterns, a moderate link-node ratio, and a high number of dominant bearing bins is labelled as \textbf{Orthogonal cities}. These cities are characterized by numerous perpendicular streets, marked by large proportion of 3-way T intersections, and has fewer street orientations. These cities also have a hierarchical road system with major arterials that can handle high volumes of traffic, while local streets serve residential and commercial areas.

Cluster 3 is characterized by a high proportion of degree 1 nodes, a significant prevalence of non-Type 1 intersections, and a low number of dominant bins. These cities display winding circuitous roads with few T and right-angled intersections. Furthermore, the low number of dominant bearing bins suggests that the street orientations are dispersed across various directions and not concentrated in any specific few. They are labelled as \textbf{Organic cities}.

\begin{table}[h]
\small
\centering
\caption{Summary of Enhanced Network Clusters}
\label{tab:enhanced_clusters}
\begin{tabular}{|p{2.5cm}|p{4cm}|p{4cm}|p{4cm}|} 
\hline
\textbf{Feature} &  \textbf{Cluster 1} & \textbf{Cluster 2}  & \textbf{Cluster 3}\\
\hline
\noalign{\smallskip}
\textbf{Node degree} & Highest prop of degree 4 nodes (4 way intersections)  & Highest prop of degree 3 nodes (3 way intersections) & Highest prop of degree 1 nodes (dead ends)  \\  
\textbf{Node pattern: Degree 3} & High proportion of Type 1 intersections & High proportion of Type 1 intersections (T intersections) & High proportion of non Type 1 intersections \\ 
\textbf{Node pattern: Degree 4} & High proportion of  Type 1   intersections & High proportion of  Type 1 and 3 intersections & No significant patterns \\
& & \\
\textbf{Link bearings} & High percentage of links are concentrated in a few specific bearing directions &  High percentage of links are concentrated in a few specific bearing directions & Links distributed in all bearing directions \\
& & \\
\textbf{Number of dominant bearing bins} & High number of dominant bearings bins (median 4) & High number of dominant bearings bins (median 4) & Few dominant bearing directions (median 0)\\
\textbf{Mean link length} & 132 & 140 & 150 \\ 
\textbf{Mean Link node ratio}  & High & Medium & Low \\ 
\textbf{Number of Cities} & 15 & 38 & 40 \\
\textbf{Example Cities} & San Francisco, Berkeley & Cupertino, Los Altos & Los Altos Hills, Walnut Creek \\
\noalign{\smallskip}
\hline
\end{tabular}
\end{table}

\begin{figure}[h]
  \includegraphics[width= 1\columnwidth]
{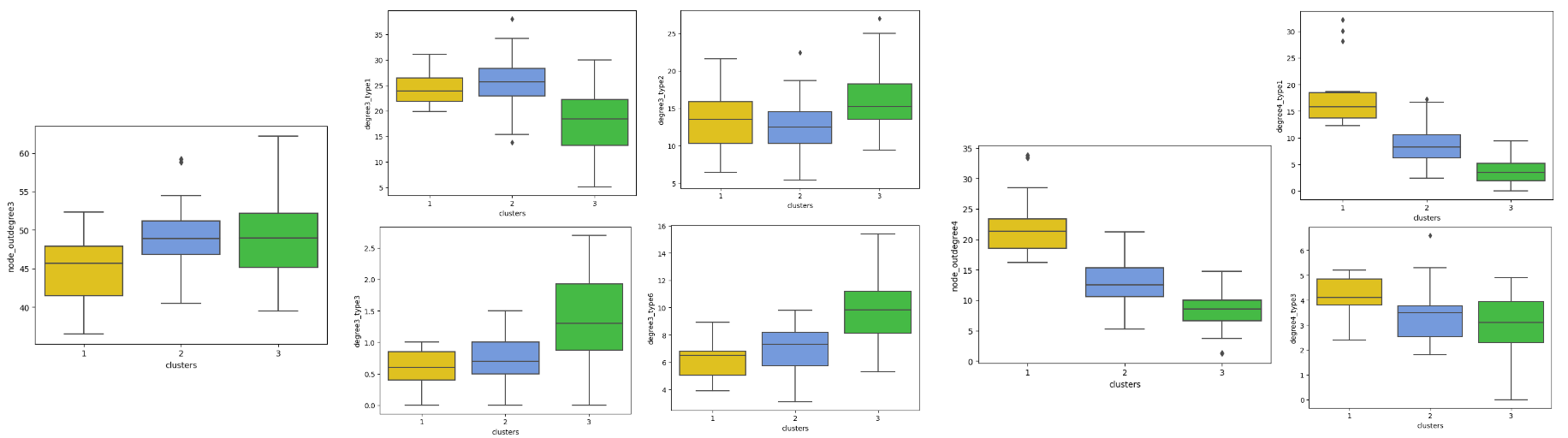}
  \caption{ The figure shows the distribution of degree 3 and 4 intersection types for enhanced clusters.}
  \label{fig:enhanced_metrics1}
\end{figure}

\begin{figure}[ht]
\includegraphics[width=.25\columnwidth]
{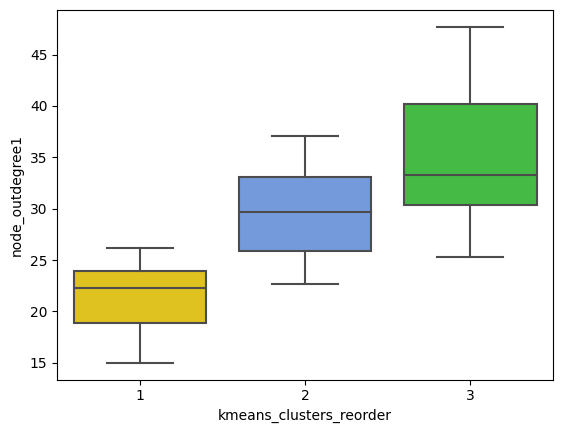}\hfill
\includegraphics[width=.25\columnwidth]
{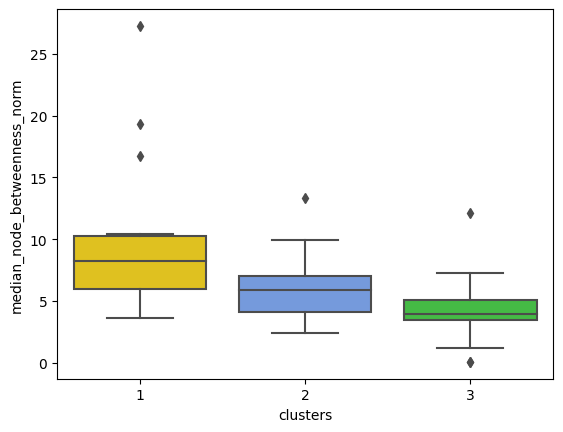}\hfill
\includegraphics[width=.25\columnwidth]
{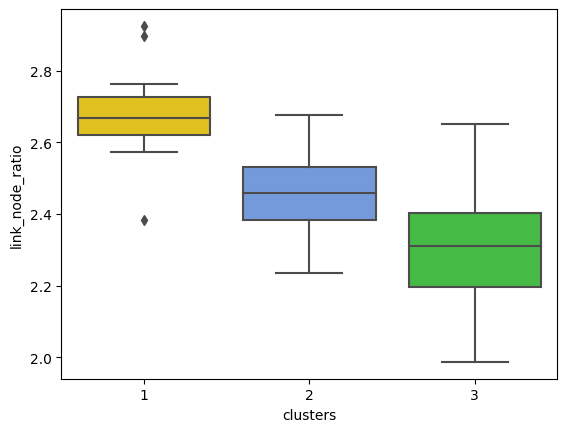}\hfill
\includegraphics[width=.25\columnwidth]
{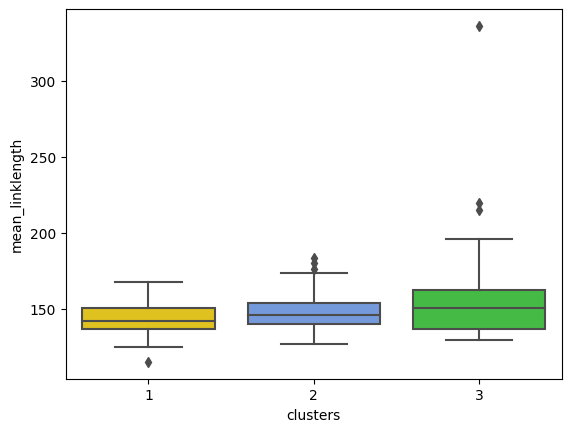}\hfill
\includegraphics[width=.25\columnwidth]
    {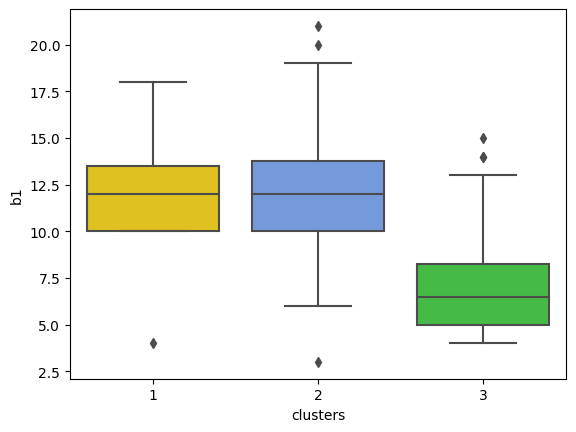}
  \includegraphics[width=.25\columnwidth]
{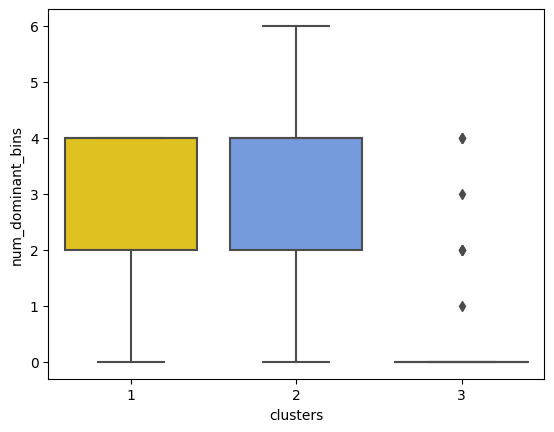}
    \caption{The figures show the distribution of relevant metrics for enhanced clusters.}
  \label{fig:enhanced_metrics2}
\end{figure}

The differentiation of node patterns for degrees 3 and 4 by clusters is illustrated using box plots in Figure \ref{fig:enhanced_metrics1}. As observed in the figure, the node types vary significantly for Orthogonal and Organic cities. Orthogonal cities exhibit a high proportion of 3-way T intersections, denoted by Type 1, whereas Organic cities have a high proportion of non-T intersections. For degree 4, the distinction is prominent for Gridded cities denoted by a high proportion of Type 1 intersections. The distribution of other important metrics for enhanced clustering results is shown in Figure \ref{fig:enhanced_metrics2}.

\section{Discussion}
\begin{figure}[ht!]
  \includegraphics[width=.5\columnwidth]
    {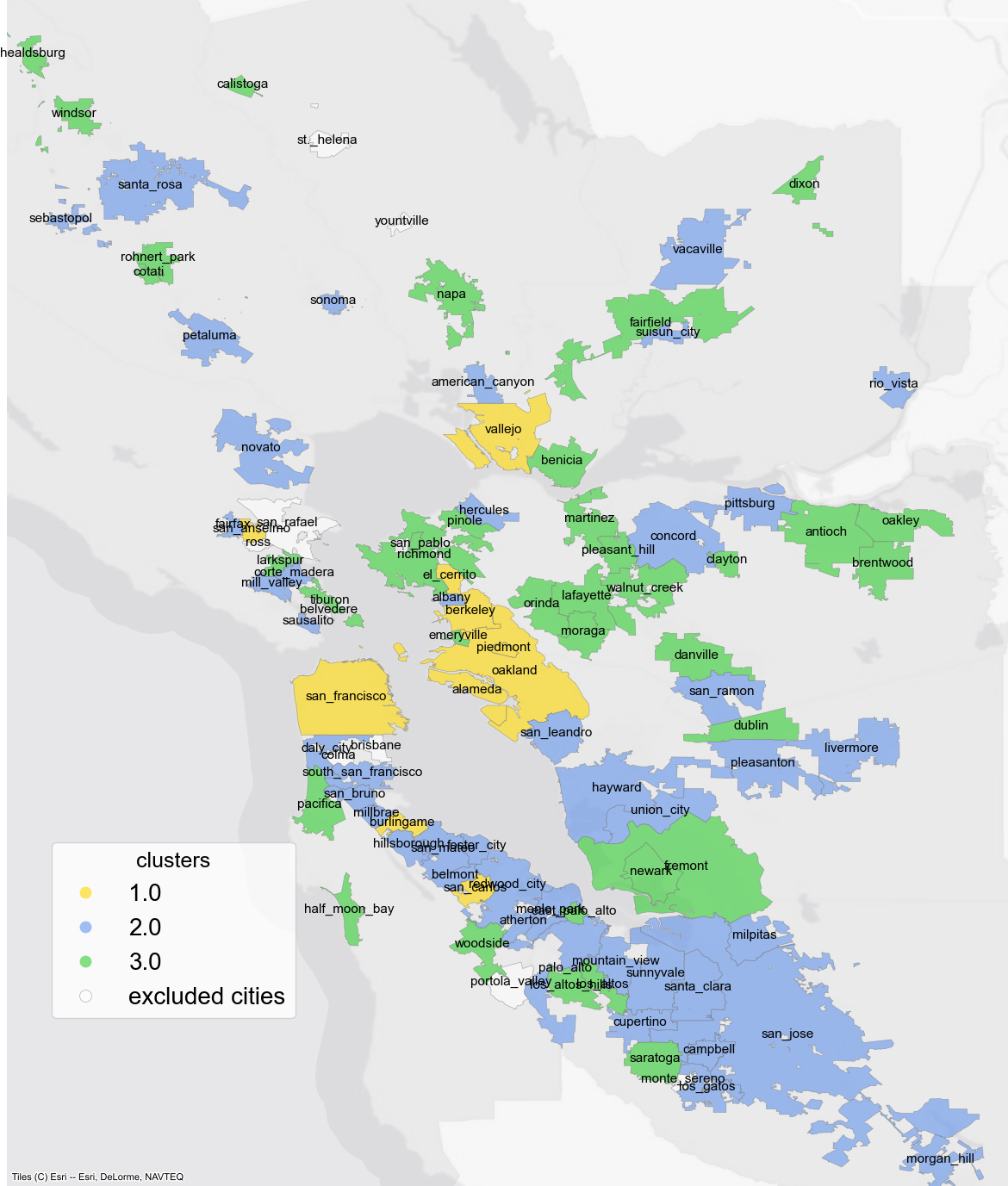}
  \includegraphics[width=.5\columnwidth]
    {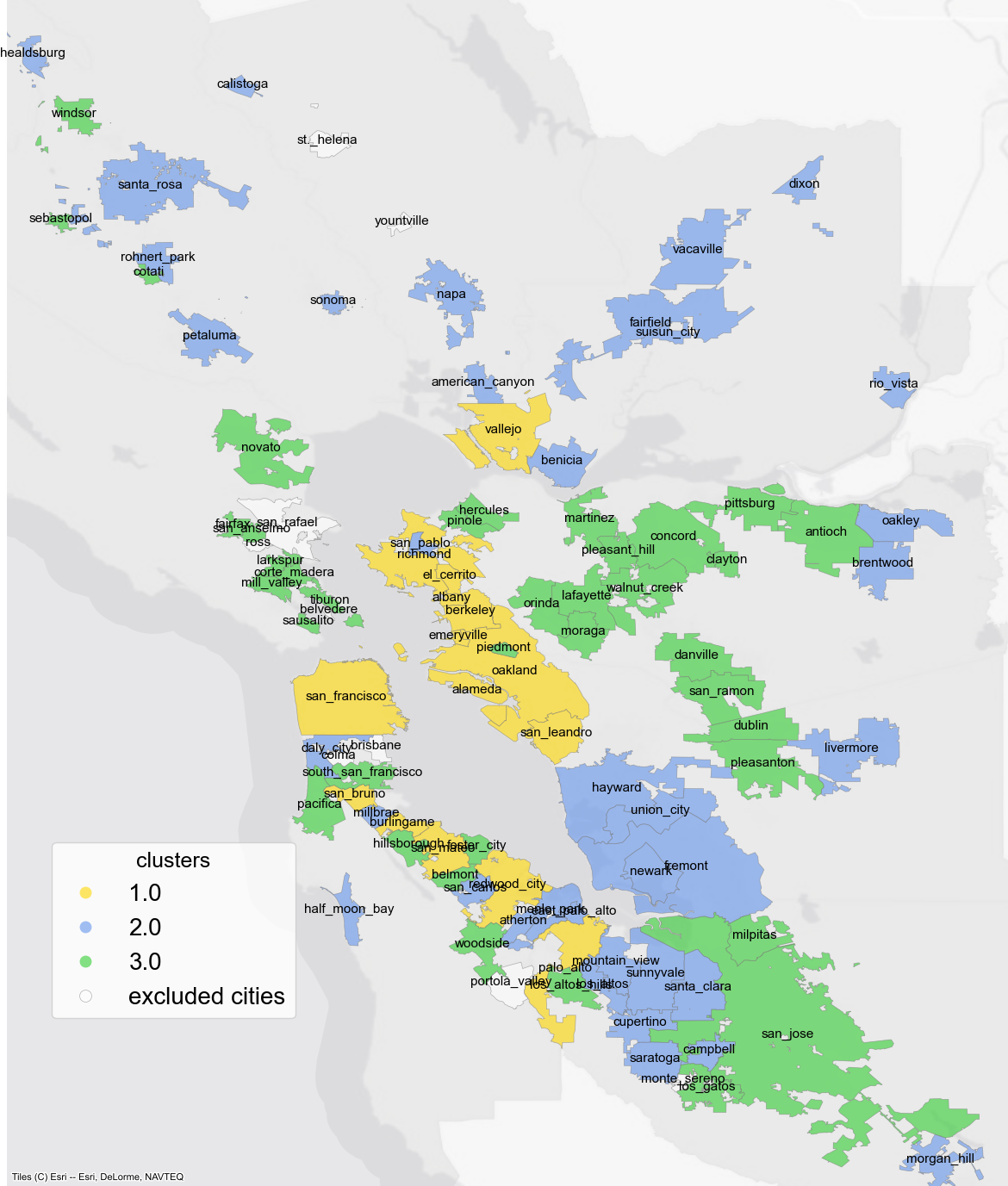}
  \caption{The figure shows the results from the baseline clustering (left) and enhanced clustering (right) for all cities in the San Francisco Bay Area.}
  \label{fig:results}
\end{figure}

In this section, we compare the cities clustered in the two approaches and identify the differences and improvements made in the enhanced method compared to the baseline. Figure \ref{fig:results} shows the typologies for all cities in the San Francisco Bay Area.

\begin{figure}[ht!]
  \includegraphics[width= 1.1\columnwidth]
{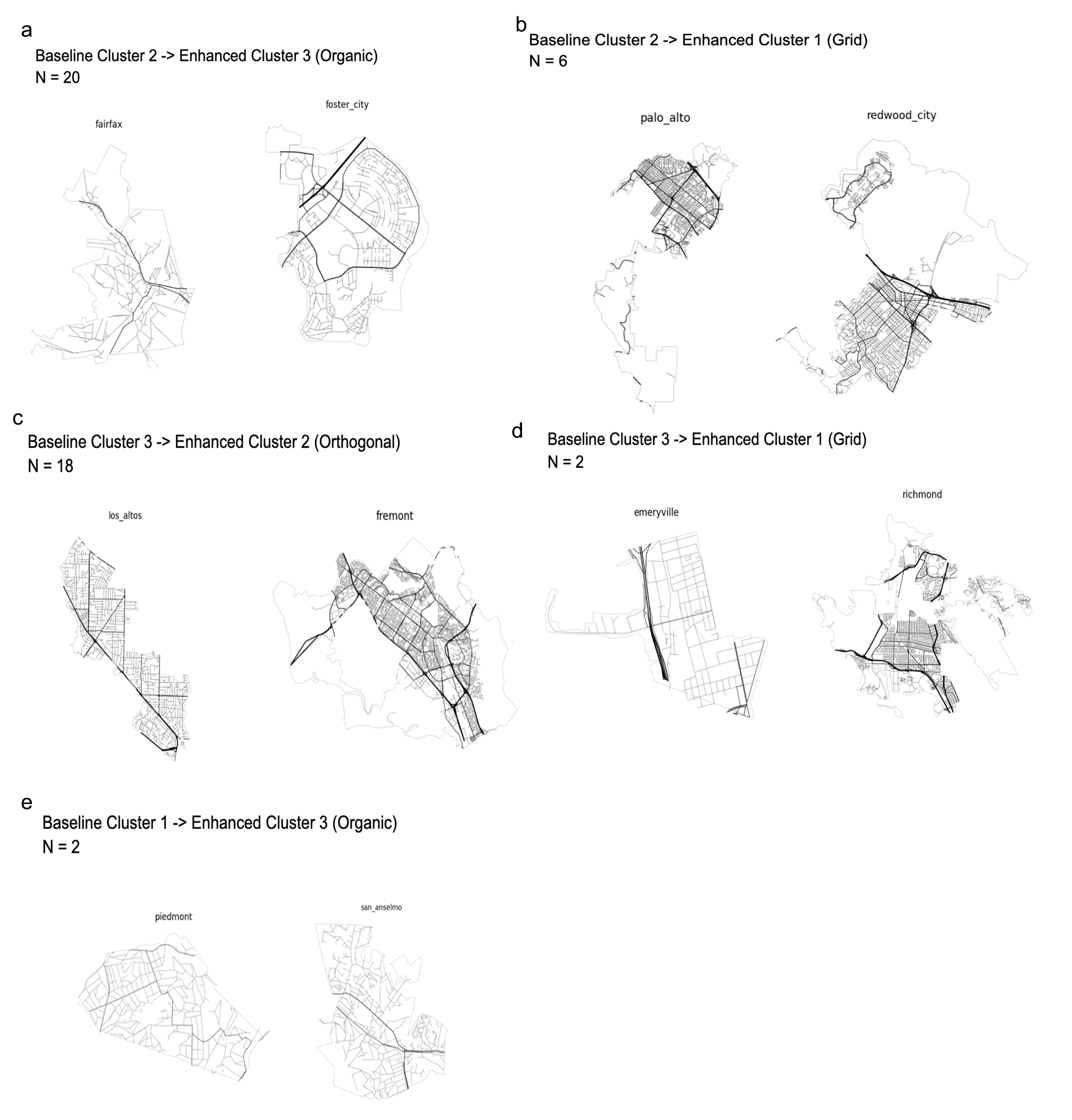}
  \caption{The figure displays the road network of some example cities that shifted between baseline and enhanced clustering. 'N' represents the total number of cities that moved.}
  \label{fig:mapping_clusters}
\end{figure}

 In the baseline clustering, Cluster 1 exhibits a high concentration of nodes with a degree of 4, making it particularly comparable to Gridded cities identified in the enhanced clustering approach. In the baseline, only 10 cities are allocated to Cluster 1. In contrast, the enhanced clustering method identifies 15 cities within the Gridded cluster. Noteworthy is the inclusion of additional cities like Albany, Emeryville, Richmond, San Mateo, etc., characterized by distinct perpendicular grid street layouts. The baseline clustering algorithm falls short of accurately capturing these cities. Additionally, in the baseline clustering, there are also two outlier cities - San Anselmo and Piedmont - that do not exhibit a gridded layout. These cities are correctly classified in the enhanced clustering (Figure \ref{fig:mapping_clusters}e).

Cluster 2 from the baseline is subdivided into two categories in the enhanced clustering. This division is determined by the geometry of the intersections: if there is a significant prevalence of right-angled T intersections, the city is placed in the Orthogonal cluster; otherwise, it is assigned to the Organic cluster. We observe that 21 cities from cluster 2 remained in the Orthogonal cluster, while 20 cities moved to the Organic cluster and 6 moved to the Gridded cluster (Figure \ref{fig:mapping_clusters}a,b). 

Based on high node degree 1, Cluster 3 in the baseline can be compared to the Organic cluster in the enhanced clustering approach. High-degree 1 cities also exhibit a substantial proportion of degree 3 nodes, forming 3-way intersections. The baseline clustering model struggles to distinguish the nuances of degree 3 intersection geometries, leading to the inclusion of cities with perpendicular 3-way intersections, such as Newark, Fremont, Los Altos, etc. This gets moved to Orthogonal in the enhanced clustering. We observe that 16 cities moved to the Orthogonal cluster, and 2 moved to the Gridded cluster(Figure \ref{fig:mapping_clusters}c,d).

Overall, for gridded cities, the primary distinction introduced by enhanced clustering lies in refining the identification of grids by explicitly considering the angles at intersections and the street bearings. Together with node degrees,  the node patterns and bearings provide a more nuanced approach. 

The most significant improvement in enhanced clustering occurs in non-gridded cities, where there is a notable prevalence of nodes with a degree of 3. While many cities have a predominance of degree 3 nodes, the specific nature of their 3-way intersections can vary significantly based on the angles involved. This variability results in either a curved layout (e.g., Los Altos Hills) or a rectangular layout (e.g., Cupertino), depending on the geometric characteristics of the intersections. The enhanced clustering with specific node patterns is adept at capturing these distinctions among cities. For example, consider Lafayette and Los Altos, both with an equal proportion of nodes with degree 3 (47\%). Despite this numerical similarity, the street and intersection geometries in these cities differ markedly. In the baseline clustering, both cities are grouped into cluster 3, overlooking their geometric nuances. However, the enhanced clustering accurately assigns Lafayette to cluster 3 and Los Altos to cluster 2. This showcases the algorithm's capability to discern and differentiate cities based on subtle variations in street and intersection patterns. By categorizing these cities into different classes, we not only differentiate them based on their road structures but also capture some essence of how drivers experience intersections upon arrival. The rectangular layout in Los Altos provides a different driving experience compared to the curved streets of Lafayette, which we believe is important to capture these nuances when classifying cities based on their network characteristics. Finally, the evaluation metrics also indicate that the enhanced clustering outperforms the baseline (refer to Table \ref{tab:evaluation_metrics}). A higher Silhouette score and a lower Davies-Bouldin Index suggest better performance for the enhanced clustering model.


\begin{table}[h]
\small
\centering
\caption{Evaluation Metrics}
\label{tab:evaluation_metrics}
\begin{tabular}{|p{4cm}|p{4cm}|p{4cm}|} 
\hline
\textbf{Metric} &  \textbf{Baseline} & \textbf{Enhanced}  \\
\hline
\noalign{\smallskip}
Silhouette Score  & 0.22  & 0.24 \\ 
Davies-Buldin Index & 1.32 & 1.16 \\
\noalign{\smallskip}
\hline
\end{tabular}
\end{table}

Classifying cities within a large urban region provides many practical insights, facilitating the development of tailored transportation policies and strategies. For instance, customizing traffic management systems can address the distinct challenges presented by gridded, organic, and orthogonal cities. One example is designing evacuation management plans tailored to accommodate three distinct types of road networks for San Francisco Bay Area. Gridded cities, characterized by well-connected, perpendicular street patterns, offer multiple entry and exit points during emergencies. This connectivity facilitates efficient movement of people and vehicles, aiding in evacuation efforts. In contrast, evacuating Organic cities, with their winding roads and dead ends, poses more challenges. The limited alternative routes underscore the need for carefully planned evacuation strategies in these cities. Orthogonal cities, known for their perpendicular street layout, require a unique approach to evacuation planning. The perpendicular nature of the streets can impede traffic flow, necessitating the prioritization of certain routes and consideration of contraflow measures. Understanding the unique characteristics of each city type allows transportation planners to devise effective evacuation strategies that minimize congestion and enhance safety.

Another potential traffic management strategy is the implementation of coordinated signal control, which can prove effective in both gridded and orthogonal cities. Conversely, organic cities, characterized by longer street lengths and winding roads, necessitate a focus on clear signage to address potential issues with dead ends and irregular intersections, thereby enhancing safety measures. 

Additionally, the formulation of new transportation policies, such as the testing of autonomous vehicles, can be tailored to specific city layouts. In the early stages, autonomous vehicles may perform better in a connected city layout like a grid with low speeds and multiple stop signs. Similarly, emerging modes of non-motorized transportation might be more suitable for a specific layout than others. Therefore, based on the network classification, a variety of policies and strategies can be designed to address the unique characteristics of each type.

Beyond its application in city clustering, the metric of intersection patterns introduced in this study can prove valuable for various purposes. Intersection angles play a crucial role in influencing traffic flow, enhancing safety measures, and contributing to overall urban design. For example, intersections characterized by acute angles may present challenges for larger vehicles, whereas those with obtuse angles may necessitate more extensive pedestrian crossings. Furthermore, optimizing the placement of traffic signals and signage can be achieved by considering the geometric characteristics of each intersection. This metric offers a versatile tool for addressing diverse aspects of urban planning and traffic management.

\section {Conclusion}

Road network plays a crucial role in shaping a city's character and influencing the quality of life for its citizens. While previous studies characterizing cities have typically focused on specific cities worldwide, there has been a lack of research concentrating on cities within large urban regions, where the street network structures can impact each other's dynamics and exhibit intricate interplay. This study addresses this gap by categorizing cities within the expansive urban region of the San Francisco Bay Area. Additionally, a new metric was developed to capture the nuances of geometry at intersections, which can better aid in classifying city networks. We conducted two clustering approaches – baseline clustering with existing metrics from the literature and enhanced clustering with additional features such as link bearings and intersection patterns.

Our findings reveal that enhanced clustering surpasses the baseline in effectively characterizing cities, identifying three distinct typologies: Gridded cities, Orthogonal cities, and Organic Cities. Gridded cities exhibit a high proportion of degree 4 nodes,  right angled 4 way intersections, 3 way T intersections, and a significant number of dominant bearing bins. Orthogonal cities are distinguished by high proportion of perpendicular streets, marked by T intersections, and fewer street orientations. Organic cities feature a high proportion of nodes with degree 1, a notable prevalence of non-Type 1 intersections, and street orientations spanning multiple directions. 

Compared to the baseline, the enhanced clustering improves the differentiation of gridded cities by explicitly considering intersection angles and street bearings, offering a more nuanced approach alongside node degrees. The most notable enhancement is observed in non-gridded cities, particularly those with a significant prevalence of nodes with degree 3. While many cities exhibit a high proportion of degree 3 nodes, the variability in intersection angles contributes to distinct layouts, whether curved or rectangular. The enhanced clustering method excels in capturing the diverse nature of their 3-way intersections, leading to more accurate classifications based on geometric characteristics.

Our study has a few limitations. The directed nature of our graph and the reliance on out-degree for node pattern identification may not fully capture the representation of intersections for one-way streets. Additionally, to enhance the analysis's accuracy in future iterations, integrating the ordering of angles into node patterns could be advantageous.

Nevertheless, this study demonstrates that incorporating nuanced geometric features enables a more realistic classification of street networks within a large urban region. When combined with network topology metrics, geometric metrics prove to be valuable tools in accurately categorizing city networks. While our current study is focused on characterizing cities based on their network structure, representing one dimension of urban environments, our future work aims to incorporate additional dimensions of transportation for a more comprehensive city characterization.


\subsection{Declaration of generative AI and AI-assisted technologies in the writing process}

During the preparation of this work the author(s) used ChatGPT in order to improve language and readability. After using this tool/service, the author(s) reviewed and edited the content as needed and take(s) full responsibility for the content of the publication.

\nocite{*}
\bibliographystyle{trb}
\bibliography{references}
\end{document}

%% file: author.tex
\begin{center}
{\Large \textbf{Beyond Centrality: Understanding Urban Street Network Typologies Through Intersection Patterns}}\\[1em]

\textbf{Anu Kuncheria, Corresponding Author}\\
\texttt{anu\_kuncheria@berkeley.edu}\\
Department of Civil and Environmental Engineering\\
University of California, Berkeley\\[0.8em]

\textbf{Joan L. Walker}\\
\texttt{joanwalker@berkeley.edu}\\
Department of Civil and Environmental Engineering\\
University of California, Berkeley\\[0.8em]

\textbf{Jane Macfarlane}\\
\texttt{janemacfarlane@berkeley.edu}\\
Smart Cities Research Center\\
University of California, Berkeley\\[1em]
\end{center}

%% file: references.bib
@article{barthelemy_self-organization_2013,
	title = {Self-organization versus top-down planning in the evolution of a city},
	volume = {3},
	copyright = {2013 The Author(s)},
	issn = {2045-2322},
	url = {https://www.nature.com/articles/srep02153},
	doi = {10.1038/srep02153},
	language = {en},
	number = {1},
	urldate = {2024-09-27},
	journal = {Scientific Reports},
	author = {Barthelemy, Marc and Bordin, Patricia and Berestycki, Henri and Gribaudi, Maurizio},
	month = jul,
	year = {2013},
	note = {Publisher: Nature Publishing Group},
	pages = {2153},
}

@article{batty_thinking_2017,
	series = {Special {Issue}: {Planning} living cities: {Patrick} {Geddes}’ legacy in the new millennium},
	title = {Thinking organic, acting civic: {The} paradox of planning for \textit{{Cities} in {Evolution}}},
	volume = {166},
	issn = {0169-2046},
	shorttitle = {Thinking organic, acting civic},
	url = {https://www.sciencedirect.com/science/article/pii/S0169204616301001},
	doi = {10.1016/j.landurbplan.2016.06.002},
	urldate = {2024-09-27},
	journal = {Landscape and Urban Planning},
	author = {Batty, Michael and Marshall, Stephen},
	month = oct,
	year = {2017},
	pages = {4--14},
}

@misc{california-law,
	title = {2023 {California} {Code} :: {Government} {Code} - {GOV} :: {TITLE} 4 - {GOVERNMENT} {OF} {CITIES} :: {DIVISION} 2 - {ORGANIZATION} {AND} {BOUNDARIES} :: {PART} 1 - {ORGANIZATION} :: {CHAPTER} 3 - {Corporate} {Name} :: {Section} 34502.},
	shorttitle = {2023 {California} {Code}},
	url = {https://law.justia.com/codes/california/code-gov/title-4/division-2/part-1/chapter-3/section-34502/},
	language = {en},
	urldate = {2024-05-16},
	journal = {Justia Law},
	month = may,
	year = {2024},
}

@article{kirkley_betweenness_2018,
	title = {From the betweenness centrality in street networks to structural invariants in random planar graphs},
	volume = {9},
	copyright = {2018 The Author(s)},
	issn = {2041-1723},
	url = {https://www.nature.com/articles/s41467-018-04978-z},
	doi = {10.1038/s41467-018-04978-z},
	language = {en},
	number = {1},
	urldate = {2023-11-20},
	journal = {Nature Communications},
	author = {Kirkley, Alec and Barbosa, Hugo and Barthelemy, Marc and Ghoshal, Gourab},
	year = {2018},
	note = {Number: 1
Publisher: Nature Publishing Group},
	pages = {2501},
}

@article{shang_statistical_2020,
	title = {Statistical {Characteristics} and {Community} {Analysis} of {Urban} {Road} {Networks}},
	volume = {2020},
	doi = {10.1155/2020/6025821},
	journal = {Complexity},
	author = {Shang, Wen-Long and Chen, Yanyan and Bi, Huibo and Zhang, Haoran and Ma, Changxi and Ochieng, Washington},
	year = {2020},
}

@article{xie_measuring_2005,
	title = {Measuring the {Structure} of {Road} {Networks}},
	volume = {39},
	doi = {10.1111/j.1538-4632.2007.00707.x},
	journal = {Geographical Analysis},
	author = {Xie, Feng and Levinson, David},
	year = {2005},
}

@article{cardillo_structural_2006,
	title = {Structural properties of planar graphs of urban street patterns},
	volume = {73},
	url = {https://link.aps.org/doi/10.1103/PhysRevE.73.066107},
	doi = {10.1103/PhysRevE.73.066107},
	number = {6},
	urldate = {2024-05-15},
	journal = {Physical Review E},
	author = {Cardillo, Alessio and Scellato, Salvatore and Latora, Vito and Porta, Sergio},
	year = {2006},
	note = {Publisher: American Physical Society},
	pages = {066107},
}

@article{wang_improved_2017,
	title = {The improved degree of urban road traffic network: {A} case study of {Xiamen}, {China}},
	volume = {469},
	shorttitle = {The improved degree of urban road traffic network},
	url = {https://ideas.repec.org//a/eee/phsmap/v469y2017icp256-264.html},
	language = {en},
	number = {C},
	urldate = {2024-05-15},
	journal = {Physica A: Statistical Mechanics and its Applications},
	author = {Wang, Shiguang and Zheng, Lili and Yu, Dexin},
	year = {2017},
	note = {Publisher: Elsevier},
	pages = {256--264},
}

@article{akbarzadeh_role_2019,
	title = {The role of travel demand and network centrality on the connectivity and resilience of an urban street system},
	volume = {46},
	issn = {1572-9435},
	url = {https://doi.org/10.1007/s11116-017-9814-y},
	doi = {10.1007/s11116-017-9814-y},
	language = {en},
	number = {4},
	urldate = {2024-05-15},
	journal = {Transportation},
	author = {Akbarzadeh, Meisam and Memarmontazerin, Soroush and Derrible, Sybil and Salehi Reihani, Sayed Farzin},
	year = {2019},
	pages = {1127--1141},
}

@article{fancello_modeling_2014,
	series = {Transportation: {Can} we do more with less resources? – 16th {Meeting} of the {Euro} {Working} {Group} on {Transportation} – {Porto} 2013},
	title = {A {Modeling} {Tool} for {Measuring} the {Performance} of {Urban} {Road} {Networks}},
	volume = {111},
	issn = {1877-0428},
	url = {https://www.sciencedirect.com/science/article/pii/S1877042814000901},
	doi = {10.1016/j.sbspro.2014.01.089},
	urldate = {2024-05-15},
	journal = {Procedia - Social and Behavioral Sciences},
	author = {Fancello, G. and Carta, M. and Fadda, P.},
	year = {2014},
	pages = {559--566},
}

@article{jiang_topological_2004,
	title = {Topological {Analysis} of {Urban} {Street} {Networks}},
	volume = {31},
	issn = {0265-8135},
	url = {https://doi.org/10.1068/b306},
	doi = {10.1068/b306},
	language = {en},
	number = {1},
	urldate = {2024-05-15},
	journal = {Environment and Planning B: Planning and Design},
	year = {2004},
	note = {Publisher: SAGE Publications Ltd STM},
	pages = {151--162},
author = {Jiang, Bin},
}

@article{boeing_urban_2019,
	title = {Urban spatial order: street network orientation, configuration, and entropy},
	volume = {4},
	issn = {2364-8228},
	shorttitle = {Urban spatial order},
	url = {https://doi.org/10.1007/s41109-019-0189-1},
	doi = {10.1007/s41109-019-0189-1},
	language = {en},
	number = {1},
	urldate = {2024-05-15},
	journal = {Applied Network Science},
	author = {Boeing, Geoff},
	year = {2019},
	pages = {67},
}

@article{tsiotas_topology_2017,
	series = {3rd {Conference} on {Sustainable} {Urban} {Mobility}, 3rd {CSUM} 2016, 26 – 27 {May} 2016, {Volos}, {Greece}},
	title = {The topology of urban road networks and its role to urban mobility},
	volume = {24},
	issn = {2352-1465},
	url = {https://www.sciencedirect.com/science/article/pii/S235214651730368X},
	doi = {10.1016/j.trpro.2017.05.087},
	urldate = {2024-05-15},
	journal = {Transportation Research Procedia},
	author = {Tsiotas, Dimitrios and Polyzos, Serafeim},
	month = jan,
	year = {2017},
	pages = {482--490},
}

@incollection{boeing_morphology_2019,
	address = {Cham},
	title = {The {Morphology} and {Circuity} of {Walkable} and {Drivable} {Street} {Networks}},
	isbn = {978-3-030-12381-9},
	url = {https://doi.org/10.1007/978-3-030-12381-9_12},
	language = {en},
	urldate = {2024-05-15},
	booktitle = {The {Mathematics} of {Urban} {Morphology}},
	publisher = {Springer International Publishing},
	author = {Boeing, Geoff},
	editor = {D'Acci, Luca},
	year = {2019},
	doi = {10.1007/978-3-030-12381-9_12},
	pages = {271--287},
}

@article{barthelemy_betweenness_2004,
	title = {Betweenness centrality in large complex networks},
	volume = {38},
	issn = {1434-6036},
	url = {https://doi.org/10.1140/epjb/e2004-00111-4},
	doi = {10.1140/epjb/e2004-00111-4},
	language = {en},
	number = {2},
	urldate = {2024-05-15},
	journal = {The European Physical Journal B},
	author = {Barthélemy, M.},
	month = mar,
	year = {2004},
	pages = {163--168},
}

@article{zhang_backbone_2017,
	title = {The backbone of urban street networks: {Degree} distribution and connectivity characteristics},
	volume = {9},
	issn = {1687-8132},
	shorttitle = {The backbone of urban street networks},
	url = {https://doi.org/10.1177/1687814017742570},
	doi = {10.1177/1687814017742570},
	number = {11},
	urldate = {2023-11-20},
	journal = {Advances in Mechanical Engineering},
	author = {Zhang, Wei and Wang, Shiguang and Tian, Xiujuan and Yu, Dexin and Yang, Zhaosheng},
	year = {2017},
	note = {Publisher: SAGE Publications},
	pages = {1687814017742570},
}

@article{masucci_exploring_2014,
	title = {Exploring the evolution of {London}'s street network in the information space: {A} dual approach},
	volume = {89},
	shorttitle = {Exploring the evolution of {London}'s street network in the information space},
	url = {https://link.aps.org/doi/10.1103/PhysRevE.89.012805},
	doi = {10.1103/PhysRevE.89.012805},
	number = {1},
	urldate = {2024-05-15},
	journal = {Physical Review E},
	author = {Masucci, A. Paolo and Stanilov, Kiril and Batty, Michael},
	month = jan,
	year = {2014},
	note = {Publisher: American Physical Society},
	pages = {012805},
}

@article{boeing_modeling_nodate,
	title = {Modeling and {Analyzing} {Urban} {Networks} and {Amenities} with {OSMnx}},
	language = {en},
	author = {Boeing, Geoff},
journal = {Wikipedia},
year = {2023},
}

@misc{noauthor_k-means_2023,
	title = {\textit{k}-means clustering},
	copyright = {Creative Commons Attribution-ShareAlike License},
	url = {https://en.wikipedia.org/w/index.php?title=K-means_clustering&oldid=1186084028},
	language = {en},
	urldate = {2023-11-24},
	journal = {Wikipedia},
	month = nov,
	year = {2023},
	note = {Page Version ID: 1186084028},
}

@article{chan_urban_2011,
	title = {Urban road networks — spatial networks with universal geometric features?: {A} case study on {Germany}’s largest cities},
	volume = {84},
	issn = {1434-6028, 1434-6036},
	shorttitle = {Urban road networks — spatial networks with universal geometric features?},
	url = {http://link.springer.com/10.1140/epjb/e2011-10889-3},
	doi = {10.1140/epjb/e2011-10889-3},
	number = {4},
	urldate = {2023-11-14},
	journal = {The European Physical Journal B},
	author = {Chan, S. H. Y. and Donner, R. V. and Lämmer, S.},
	month = dec,
	year = {2011},
	pages = {563--577},
}

@article{burghardt_road_2022,
	title = {Road network evolution in the urban and rural {United} {States} since 1900},
	volume = {95},
	issn = {0198-9715},
	url = {https://www.sciencedirect.com/science/article/pii/S0198971522000473},
	doi = {10.1016/j.compenvurbsys.2022.101803},
	journal = {Computers, Environment and Urban Systems},
	author = {Burghardt, Keith and Uhl, Johannes H. and Lerman, Kristina and Leyk, Stefan},
	month = jul,
	year = {2022},
	pages = {101803},
}

@article{dumedah_characterising_2021,
	title = {Characterising the structural pattern of urban road networks in {Ghana} using geometric and topological measures},
	volume = {8},
	issn = {2054-4049},
	url = {https://onlinelibrary.wiley.com/doi/abs/10.1002/geo2.95},
	doi = {10.1002/geo2.95},
	number = {1},
	urldate = {2023-11-16},
	journal = {Geo: Geography and Environment},
	author = {Dumedah, Gift and Garsonu, Emmanuel Kofi},
	year = {2021},
	note = {\_eprint: https://onlinelibrary.wiley.com/doi/pdf/10.1002/geo2.95},
	pages = {e00095},
}

@misc{noauthor_osmnx_2016,
	title = {{OSMnx}: {Python} for {Street} {Networks}},
	shorttitle = {{OSMnx}},
	url = {https://geoffboeing.com/2016/11/osmnx-python-street-networks/},
	language = {en-US},
	urldate = {2023-11-24},
	journal = {Geoff Boeing},
	month = nov,
	year = {2016},
}

@misc{mtc,
  title = {San {Francisco} {Bay} {Region} {Incorporated} {Cities} and {Towns}},
  howpublished = {\url{https://opendata.mtc.ca.gov/datasets/MTC::san-francisco-bay-region-incorporated-cities-and-towns-1/about},
  note = {Accessed: 2022-02-20}},
year = {2023},
}

@article{chan_simulating_2022,
	title = {Simulating the {Impact} of {Dynamic} {Rerouting} on {Metropolitan}-scale {Traffic} {Systems}},
	volume = {33},
	issn = {1049-3301},
	url = {https://doi.org/10.1145/3579842},
	doi = {10.1145/3579842},
	number = {1-2},
	urldate = {2023-03-01},
	journal = {ACM Transactions on Modeling and Computer Simulation},
	author = {Chan, Cy and Kuncheria, Anu and Macfarlane, Jane},
	year = {2023},
	pages = {7:1--7:29},
}

@misc{here_tech,
        author = {{HERE Technologies}},
        howpublished = "\url{https://www.here.com/}",
        year = {2019},
        note = "[Online; accessed 06-Feb-2019]"
}

@article{buhl_topological_2006,
	title = {Topological patterns in street networks of self-organized urban settlements},
	volume = {49},
	doi = {10.1140/epjb/e2006-00085-1},
	journal = {Eur Phys J B},
	author = {Buhl, J. and Gautrais, Jacques and Reeves, N. and Sole, Ricard and Valverde, Sergi and Kuntz, Pascale and Theraulaz, Guy},
	month = feb,
	year = {2006},
	pages = {513--522},
}

@article{jiang_topological_2007,
	title = {A topological pattern of urban street networks: {Universality} and peculiarity},
	volume = {384},
	issn = {0378-4371},
	shorttitle = {A topological pattern of urban street networks},
	url = {https://www.sciencedirect.com/science/article/pii/S0378437107006140},
	doi = {10.1016/j.physa.2007.05.064},
	urldate = {2023-11-24},
	journal = {Physica A: Statistical Mechanics and its Applications},
	author = {Jiang, Bin},
	month = oct,
	year = {2007},
	pages = {647--655},
}

@article{xie_measuring_2007,
	title = {Measuring the {Structure} of {Road} {Networks}},
	volume = {39},
	issn = {1538-4632},
	url = {https://onlinelibrary.wiley.com/doi/abs/10.1111/j.1538-4632.2007.00707.x},
	doi = {10.1111/j.1538-4632.2007.00707.x},
	number = {3},
	urldate = {2023-11-10},
	journal = {Geographical Analysis},
	author = {Xie, Feng and Levinson, David},
	year = {2007},
	note = {\_eprint: https://onlinelibrary.wiley.com/doi/pdf/10.1111/j.1538-4632.2007.00707.x},
	pages = {336--356},
}

@article{badhrudeen_geometric_2022,
	title = {A {Geometric} {Classification} of {World} {Urban} {Road} {Networks}},
	volume = {6},
	issn = {2413-8851},
	url = {https://www.mdpi.com/2413-8851/6/1/11},
	doi = {10.3390/urbansci6010011},
	language = {en},
	number = {1},
	urldate = {2022-09-29},
	journal = {Urban Science},
	author = {Badhrudeen, Mohamed and Derrible, Sybil and Verma, Trivik and Kermanshah, Amirhassan and Furno, Angelo},
	month = feb,
	year = {2022},
	pages = {11},
}

@article{strano_urban_2013,
	title = {Urban {Street} {Networks}, a {Comparative} {Analysis} of {Ten} {European} {Cities}},
	volume = {40},
	issn = {0265-8135, 1472-3417},
	url = {http://journals.sagepub.com/doi/10.1068/b38216},
	doi = {10.1068/b38216},
	language = {en},
	number = {6},
	urldate = {2023-11-14},
	journal = {Environment and Planning B: Planning and Design},
	author = {Strano, Emanuele and Viana, Matheus and Da Fontoura Costa, Luciano and Cardillo, Alessio and Porta, Sergio and Latora, Vito},
	month = dec,
	year = {2013},
	pages = {1071--1086},
}

@article{crucitti_centrality_2006,
	title = {Centrality measures in spatial networks of urban streets},
	volume = {73},
	issn = {1539-3755, 1550-2376},
	url = {https://link.aps.org/doi/10.1103/PhysRevE.73.036125},
	doi = {10.1103/PhysRevE.73.036125},
	language = {en},
	number = {3},
	urldate = {2023-11-24},
	journal = {Physical Review E},
	author = {Crucitti, Paolo and Latora, Vito and Porta, Sergio},
	month = mar,
	year = {2006},
	pages = {036125},
}

@article{lin_comparative_2017,
	title = {Comparative {Analysis} on {Topological} {Structures} of {Urban} {Street} {Networks}},
	volume = {6},
	copyright = {http://creativecommons.org/licenses/by/3.0/},
	issn = {2220-9964},
	url = {https://www.mdpi.com/2220-9964/6/10/295},
	doi = {10.3390/ijgi6100295},
	number = {10},
	urldate = {2023-11-17},
	journal = {ISPRS International Journal of Geo-Information},
	author = {Lin, Jingyi and Ban, Yifang},
	month = oct,
	year = {2017},
	note = {Number: 10
Publisher: Multidisciplinary Digital Publishing Institute},
	pages = {295},
}

@article{popovich_methodology_2021,
	title = {A methodology to develop a geospatial transportation typology},
	volume = {93},
	issn = {0966-6923},
	url = {https://www.sciencedirect.com/science/article/pii/S0966692321001149},
	doi = {10.1016/j.jtrangeo.2021.103061},
	language = {en},
	urldate = {2023-05-30},
	journal = {Journal of Transport Geography},
	author = {Popovich, Natalie and Spurlock, C. Anna and Needell, Zachary and Jin, Ling and Wenzel, Tom and Sheppard, Colin and Asudegi, Mona},
	month = may,
	year = {2021},
	pages = {103061},
}

@article{oke_novel_2019,
	title = {A novel global urban typology framework for sustainable mobility futures},
	volume = {14},
	issn = {1748-9326},
	url = {https://dx.doi.org/10.1088/1748-9326/ab22c7},
	doi = {10.1088/1748-9326/ab22c7},
	language = {en},
	number = {9},
	urldate = {2023-05-30},
	journal = {Environmental Research Letters},
	author = {Oke, Jimi B. and Aboutaleb, Youssef M. and Akkinepally, Arun and Azevedo, Carlos Lima and Han, Yafei and Zegras, P. Christopher and Ferreira, Joseph and Ben-Akiva, Moshe E.},
	month = sep,
	year = {2019},
	note = {Publisher: IOP Publishing},
	pages = {095006},
}
